\documentclass{aa}
\usepackage[varg]{txfonts}
\usepackage{morefloats}
\usepackage{color}
\usepackage{amsmath}
\usepackage{multirow}
\usepackage{graphicx}

\newcommand{\msun}{\ensuremath{M_{\odot}}}

\newcommand{\HI}{H\,{\sc i}{}}
\begin{document}
\title{A low \HI\ column density filament in  NGC 2403:\\ signature of interaction or accretion} 
\author{W.J.G. de
  Blok\inst{1,2,3}, Katie M.\ Keating\inst{4},
  D.J. Pisano\inst{5,6}, F. Fraternali\inst{7,3},
  F. Walter\inst{8}, T. Oosterloo\inst{1,3},
  E. Brinks\inst{9}, F. Bigiel\inst{10},
  A. Leroy\inst{11}} 
\institute{Netherlands Institute
  for Radio Astronomy (ASTRON), Postbus 2, 7990 AA Dwingeloo, the
  Netherlands, {\tt blok@astron.nl}
\and
Astrophysics,
  Cosmology and Gravity Centre, Department of Astronomy, University of
  Cape Town, Private Bag X3, Rondebosch 7701, South Africa
\and
Kapteyn Astronomical Institute, University of
  Groningen, PO Box 800, 9700 AV Groningen, the Netherlands
\and
Rincon Research Corporation, Tucson, AZ 85711, USA
\and
WVU Dept.\ of Physics \& Astronomy, P.O. Box 6315,
  Morgantown, WV 26506, USA
\and
Adjunct Assistant
  Astronomer at NRAO-Green Bank, P.O. Box 2, Green Bank, WV 24944,
  USA
\and
Department of Physics and Astronomy,
  University of Bologna, via Berti Pichat 6/2, I-40127 Bologna, Italy
\and
Max-Planck Institut f\"ur Astronomie, K\"onigstuhl
  17, 69117 Heidelberg, Germany
\and
Centre for Astrophysics Research, University of
  Hertfordshire, College Lane, Hatfield AL10 9AB, United Kingdom
\and
Institut f\"ur theoretische Astrophysik, Zentrum
  f\"ur Astronomie der Universit\"at Heidelberg, Albert-Ueberle
  Str.\ 2, 69120 Heidelberg, Germany
\and
National Radio Astronomy Observatory, 520 Edgemont Road, Charlottesville, VA 22903, USA}

\date{Received 1 January 1900 / Accepted 1 January 1900}

\abstract{Observed \HI\ accretion around nearby galaxies can only account for a
fraction of the gas supply needed to sustain the currently observed star
formation rates. It is possible that additional accretion happens in the form of
low column density cold flows, as predicted by numerical simulations of
galaxy formation. To contrain the presence and properties of such
flows, we present deep \HI\ observations obtained with the NRAO Green
Bank Telescope of an area measuring $4^{\circ} \times 4^{\circ}$
around NGC 2403.  These observations, with a 5$\sigma$ detection limit
of $2.4 \cdot 10^{18}$ cm$^{-2}$ over a 20 km s$^{-1}$ linewidth,
reveal the presence of a low-column density, extended cloud outside
the main \HI\ disk, about $17'$ ($\sim 16$ kpc or $\sim 2\, R_{25}$)
to the NW of the center of the galaxy. The total \HI\ mass of the
cloud is $6.3 \cdot 10^6\ \msun$, or 0.15 percent of the total
\HI\ mass of NGC 2403. The cloud is associated with an 8-kpc
anomalous-velocity \HI\ filament in the inner disk, previously
observed in deep VLA observations by Fraternali et al.\ (2001,
2002). We discuss several scenarios for the origin of the cloud, and
conclude that it is either accreting from the intergalactic medium,
or  is the result of a minor interaction with a
neigbouring dwarf galaxy.}


\titlerunning{Low column density \HI\ near NGC 2403}
\authorrunning{de Blok et al.}
\maketitle

\section{Introduction}

Accretion of gas from the intergalactic medium (IGM) is often proposed
as a mechanism for galaxies to obtain sufficient gas to maintain their
star formation rate over cosmic time. There is however little direct
observational evidence that can constrain the existence of the
accretion and the magnitude of the accretion rates. In a recent
review, \citet{sancisi08} make an inventory of possible signatures of
gas accretion in and around nearby galaxies.  They find an average
visible \HI\ accretion rate of $\sim 0.2\ M_{\odot}$ year$^{-1}$,
which is an order of magnitude too low to sustain the current star
formation rates of the galaxies studied.

Searches for accreting gas are hampered by the difficulty of
disentangling external, accreting gas from gas that has been expelled
from the disk by processes of star formation (the so-called ``galactic
fountain''; \citealt{shapiro}). In many cases, sophisticated
three-dimensional modeling of the \HI\ data cube is required to do
this. Recent examples of this kind of analysis are presented in papers
from the HALOGAS survey (Hydrogen Accretion in LOcal GAlaxieS;
\citealt{heald11}), a deep \HI\ survey of 24 nearby galaxies, looking,
amongst others, for signs of accretion in these galaxies (see
\citealt{hg1, hg2, hg3, hg4}).

These issues were also highlighted in earlier deep surveys of two
nearby galaxies, NGC 891 and NGC 2403.  Deep \HI\ observations of the
edge-on galaxy NGC 891 showed the presence of \HI\ far above the main
disk \citep{swaters97, oosterloo07}. Most of this \HI\ was found to be
rotating with the disk, but lagging slightly. This slower rotation of
the extra-planar gas can be explained within the context of the
galactic fountain model \citep{ff_binney06}. A smaller fraction of the
extra-planar \HI\ in NGC 891 was found to be at velocities
inconsistent with circular rotation \citep{oosterloo07}. This gas is
thought to be accreting. A number of the prominent \HI\ filaments in
NGC 891 are also difficult to explain with the galactic fountain
model.

A second galaxy where prominent, lagging extra-planar \HI\ is
detected is NGC 2403 \citep{ff01,ff02}. This galaxy is observed at an
intermediate inclination of $63^{\circ}$, which has the advantage that
signs of in- or outflow of the gas can be detected directly. Modeling
by  \citet{ff01,ff02} and \citet{ff_binney08} shows that the
kinematics of the extraplanar gas in NGC 2403 is consistent with
inflow and this could therefore be a sign of accretion.

The analyses described above were done using data obtained with the
Very Large Array (VLA) or the Westerbork Synthesis Radio Telescope
(WSRT), i.e., radio interferometric telescopes. While these have
excellent spatial resolution, a large observing effort is required to
reach the low column density levels where one expects to see signs of
accretion.  For this reason, we have started a survey of a number of
nearby galaxies (mainly taken from the THINGS survey;
\citealt{walter08}) using the Green Bank Telescope (GBT), with the
goal to obtain the deepest maps of these galaxies obtained so far, and
quantify the presence of low-column density, and possibly accreting,
features around these galaxies. A full description and first results
of this survey will be presented elsewhere (\citealt{pisano2014},
Pisano et al., in prep.).

Here we report on the discovery, using GBT observations, of an
extended, low-column density \HI\ feature close to and possibly
connected with NGC 2403.  We describe the observations in Sect.\ 2.
In Sect.\ 3, we argue that the feature is associated with an
anomalous-velocity \HI\ complex discovered in earlier high-resolution
VLA observations \citep{ff01,ff02}. In Sect.\ 4 we describe various
evolution scenarios for the feature.  We summarize our results in
Sect.\ 5.

\section{Observations}

NGC 2403 is a nearby ($D=3.2$ Mpc; $1' = 0.93$ kpc) late-type spiral,
and part of the M81 group.  It has an absolute luminosity $M_B =
-19.4$, an extended \HI\ disk, and an inclination of $63^{\circ}$. It
has a well-defined, symmetrical rotation curve, with a maximum
rotation velocity of 135 km s$^{-1}$.  Its heliocentric systemic
velocity is 132.8 km s$^{-1}$. See \citet{walter08, deblok08} and
\citet{trachternach08} for more detailed information regarding 
general properties of NGC 2403.

\subsection{GBT observations}

Observations of NGC 2403 were carried out with the 100m Robert
C.\ Byrd Green Bank Telescope (GBT) of the NRAO\footnote{The National
  Radio Astronomy Observatory (NRAO) is a facility of the National
  Science Foundation operated under cooperative agreement by
  Associated Universities, Inc.}  in 21 sessions between 29 May and 30
September 2009.  We observed a $4^{\circ} \times 4^{\circ}$ area
centered on NGC 2403.  Strips of constant right ascension and
declination, spaced by $3'$, were observed to form a `basket weave'
pattern over the region.  The spacing and integration times ensure
Nyquist-sampling. The total integration time for the entire map was
approximately 127 hours, corresponding to an integration time per beam
of $\sim 11$ min. The GBT spectrometer was used with a bandwidth of
12.5 MHz. This corresponds to a velocity range from $-885$ km s$^{-1}
< v_{\rm LSR} < 1750$ km s$^{-1}$. The typical system temperature for each
channel of the dual-polarization receiver was $\sim$ 20 K.

A reference spectrum for each of the sessions was made by
observing an emission-free region about $4.75^{\circ}$ away from the
galaxy. The reference spectrum was then used to perform a
(signal--reference)/reference calibration of each pixel.  The
calibrated spectra were scaled by the system temperature, corrected
for atmospheric opacity and GBT efficiency. We adopted the GBT
efficiency equation~(1) from \citet{lang07} with a zenith atmospheric
opacity $\tau_{0}$ = 0.009. Data reduction was done using custom
routines developed using GBTIDL\footnote{http://gbtidl.nrao.edu}.

The frequency range observed was relatively free of radio-frequency
interference (RFI), with $\sim 5\%$ of all spectra adversely
affected. Spectra that showed high noise values across many channels
were flagged and removed. After amplitude calibration and gridding, a
first-order polynomial was fit to line-free regions of the spectra and
subtracted from the gridded spectra.

Data from all observing sessions were converted to heliocentric
velocities and combined into a single data cube.  In order to match
the typical linewidths of neutral hydrogen features observed around
galaxies, spectra were smoothed to a velocity resolution of 5.2 km
s$^{-1}$.  The effective angular resolution is $8.7'$, and we use a
pixel size of $1.75'$. The calibration from K to Jy was derived by
observing 3C286 in the same way as the \HI\ maps. The final scale
factor from K to Jy is $0.43 \pm 0.03$. Due to remaining low-level RFI
and residual continuum emission, the noise level varies by $\sim 20\%$
throughout the cube. The average RMS noise in the final data cube is
$6.0$ mJy beam$^{-1}$ or 13.6 mK per 5.2 km s$^{-1}$ channel. This
translates to a $1\sigma$ column density sensitivity of $1.3 \cdot
10^{17}$ cm$^{-2}$ per channel, or a $5\sigma$ sensitivity of
$2.4\cdot 10^{18}$ cm$^{-2}$ for a 20 km s$^{-1}$ full width at half
maximum \HI\ line.

\subsection{Deep VLA observations\label{sec:vladata}}

In this paper we compare the GBT data with deep VLA observations
obtained by \citet{ff01,ff02}. They observed NGC 2403 for 40 hours
with the VLA in its CS configuration.  Here we use the $30''$ data
presented in \citet{ff02}. The data cube was created using a
robustness parameter of 0 and a taper of $27''$. The channel spacing
is 5.15 km s$^{-1}$. The data were Hanning-smoothed, leading to an
effective velocity resolution of 10.3 km s$^{-1}$. The synthesized
beam size of this cube is $29.7'' \times 29.3''$ and the column
density sensitivity limit for a $5\sigma$ detection over one velocity
resolution element is $2.0 \cdot 10^{19}$ cm$^{-2}$.  See \citet{ff02}
for more details.

In addition, we also smoothed the $30''$ cube spatially to a
resolution of $60''$ and $90''$, as well as to the GBT resolution of
8.7$'$ (using a simple Gaussian smoothing function). We use these
cubes later to determine how much of the \HI\ detected with the GBT
was also detected by the deep VLA observations.

\section{Analysis}

\subsection{GBT channel maps and moment maps}

To derive the total \HI\ mass and the global velocity profile, we
clipped the GBT data cube at 2.5$\sigma$, also removing by hand a small number
of spurious noise peaks (caused by
low-level residual RFI). The rotation and systemic velocity of NGC
2403 mean that for radial velocities below $v_{\rm hel} \sim 10$ km
s$^{-1}$ galaxy emission overlaps with Galactic emission. Based on
symmetry considerations the velocity range where we cannot recover the
NGC 2403 \HI\ distribution is however small.

\begin{figure}
\centerline{\includegraphics[width=0.45\textwidth]{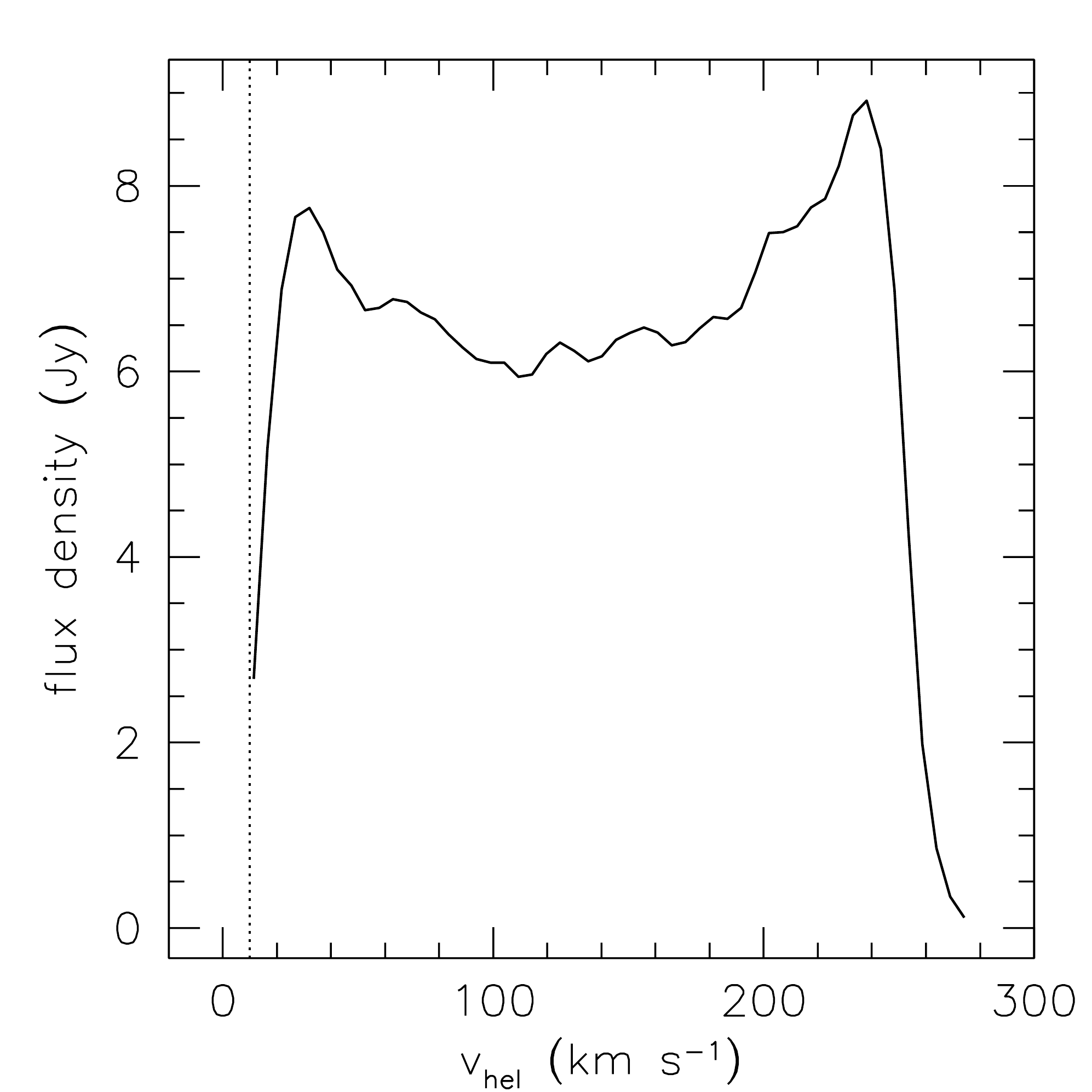}}
\caption{Global \HI\ profile of NGC 2403 derived from the GBT
  data. The dotted line indicates $v_{\rm hel} = 10$ km s$^{-1}$.
  Below this velocity, Galactic emission affects the data and emission
  from NGC 2403 could not be unambiguously
  identified.\label{fig:globprof}}
\end{figure}

Figure \ref{fig:globprof} shows the global \HI\ profile. We find a
total flux of 1677.1 Jy km s$^{-1}$. This does not include any
emission below $v_{\rm hel} = 10$ km s$^{-1}$, but it is clear from
Fig.~\ref{fig:globprof} that this will be only a small fraction of the
total flux.  Assuming a distance of 3.2 Mpc, we find an \HI\ mass of
$4.03 \cdot 10^{9}$ \msun. This is higher than was found in previous
observations.  Single-dish data by \citet{rots80} yield a total flux
of 1448.1 Jy km s$^{-1}$ or an \HI\ mass of $3.50 \cdot 10^9$ \msun.
Interferometric data by \citet{begeman87} gave an \HI\ mass of $3.0
\cdot 10^9$ \msun, observations by \citet{sicking97} $3.13 \cdot 10^9$
\msun, and deep observations by \citet{ff02} $3.28 \cdot 10^9$
\msun\ (all values corrected to $D=3.2$ Mpc). 

Uncertainties in interferometrically measured \HI\ masses are
  usually dominated by calibration uncertainties, bandpass
  corrections, selection of masks, etc., rather than by the formal
  statistical uncertainties. They are therefore difficult to
  quantify. An empirically motivated estimate of the uncertainties is
  $\sim 10\%$. This value is higher than the formal statistical
  uncertainty, which tends to be a few percent at most for bright
  sources such as considered here (cf.\ the values quoted in
  \citealt{begeman87} and \citealt{sicking97}). Comparing the above
  measurements with the GBT observations, we see that the latter
  detect 20\% more flux, i.e., a significantly higher value.

Part of this additional flux will be due to the larger area observed
by the GBT compared to the single interferometer pointings.  However,
we also detect more flux than found in the \citet{rots80} single-dish
data. \citet{rots80} surveyed a similar area as our observations, so
the difference cannot be due to a more limited field of view.  A comparison
between the two single-dish global profiles, shows that our
observations detect more flux over the entire velocity range of the
galaxy. Although this could be a calibration issue, it is likely that
our new GBT observations, which are approximately 50--100 times more
sensitive in terms of column density, detect more low column density
\HI. For example, the low-column density kinematically anomolous
\HI\ detected by \citet{ff01,ff02} has a total \HI\ mass of $\sim 3
\cdot 10^8\, \msun$, which could explain a large part of the
difference in the derived single-dish masses.

\begin{figure*}
\centerline{\includegraphics[width=0.95\textwidth]{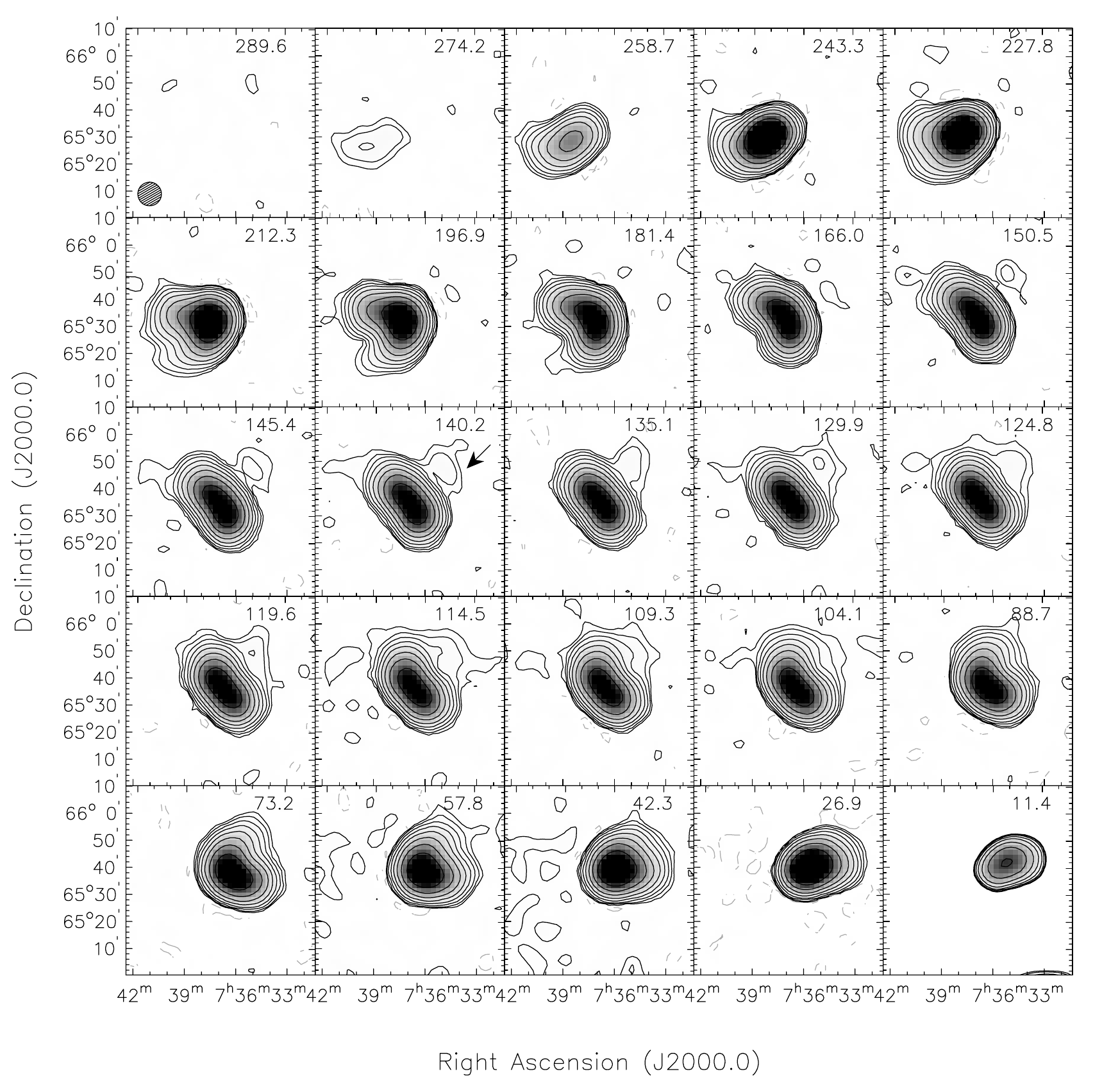}}
\caption{Selected channel maps of the GBT data cube. Contours levels
  are $-2 \sigma$ (gray dashed contours) and
  $(2,4,8,16,32,64,128,256,512,1024) \cdot \sigma$ (full contours)
  where $\sigma=6$ mJy beam$^{-1}$. This corresponds within a single
  channel to a column density of $1.25 \cdot 10^{17}$
  cm$^{-2}$. Heliocentric velocities of the channels in km s$^{-1}$
  are shown in the top-right of each channel.  From 150.5 to 104.1 km
  s$^{-1}$ every channel is shown. Outside this range only every third
  channel is shown. The GBT beam is shown in the bottom-left corner of
  the top-left panel. Channels in the bottom-row are marginally
  affected by Galactic emission. In the 140.2 km s$^{-1}$ panel the
  arrow indicates cloud emission.
\label{fig:chans}}
\end{figure*}

Selected channel maps from the GBT data cube are shown in
Fig.~\ref{fig:chans}. The emission from the main disk of NGC 2403 is
symmetric and regular, though the velocity field (see below) indicates
that the outer disk is slightly warped.  In the channel maps between
$v=150.5$ and $104.1$ km s$^{-1}$ we see a faint extension to the NW
of the main emission component, indicated by the  arrow in
the $v=140.2$ km s$^{-1}$ panel. This ``cloud'' seems to connect up
with the main disk at lower velocities, but looks separate from it
towards higher velocities.
All velocities mentioned here and below are
heliocentric velocities.

\begin{figure*}
\centering{\includegraphics[width=0.8\textwidth]{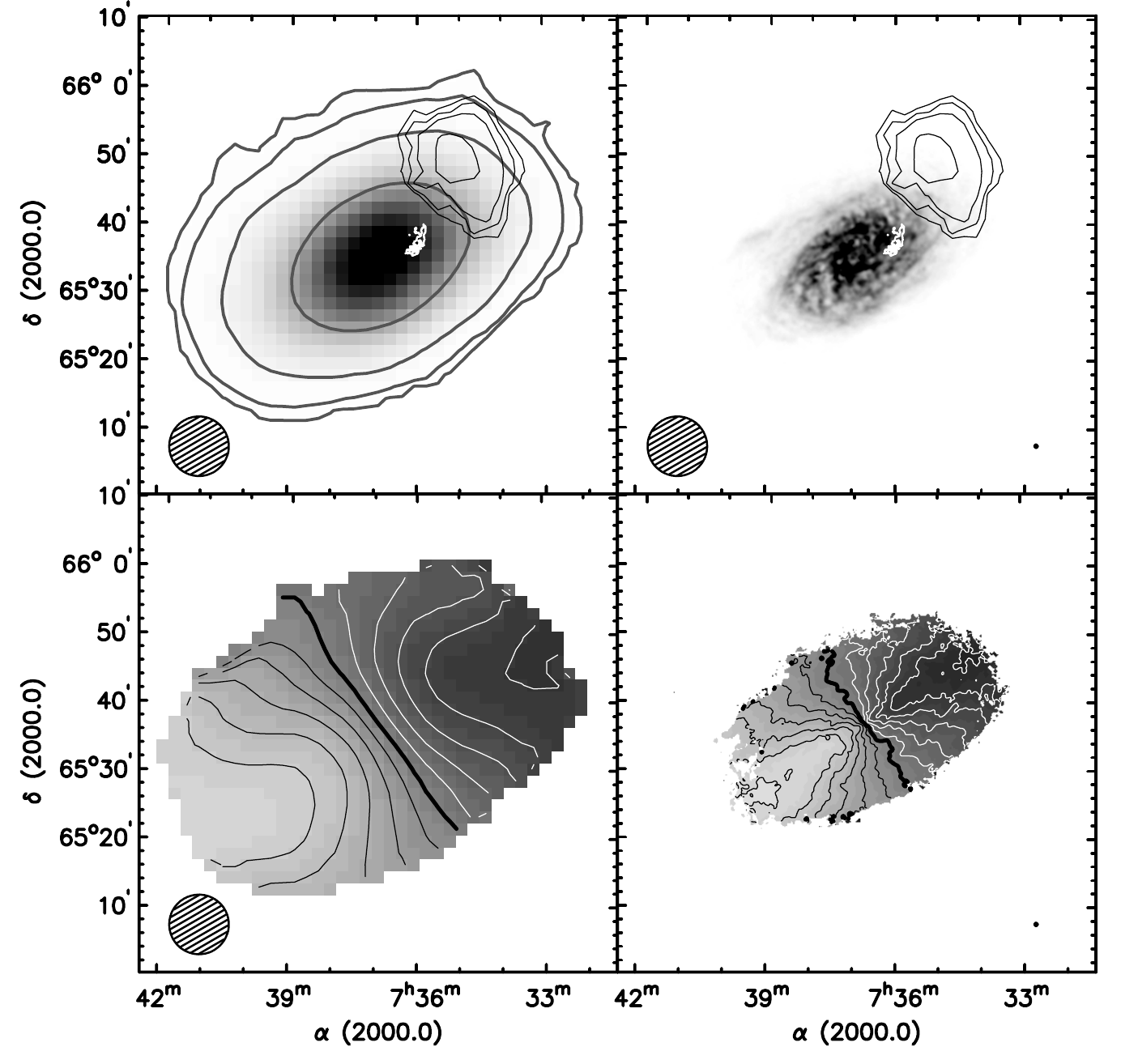}}
\caption{\HI\ moment maps of NGC 2403. {\bf Top-left:} Integrated
  \HI\ map derived from the GBT data. Thick dark-gray contours show
  the \HI\ distribution of the entire galaxy. Contour levels are
  $5\cdot 10^{17},\, 5\cdot 10^{18},\, 5\cdot 10^{19},\ {\rm
    and}\ 5\cdot 10^{20}$ cm$^{-2}$. The gray-scale runs from $5 \cdot
  10^{17}$ (white) to $1.5 \cdot 10^{21}$ cm$^{-2}$ (black). Thin, black
  contours show the \HI\ distribution of the cloud. Contour levels are
  $(6.25,\, 12.5,\, 25,\, 62.5)\cdot 10^{17}$ cm$^{-2}$. White
  contours indicate the location of the 8-kpc filament. The GBT beam
  is shown in the bottom-left. {\bf Top-right:} Integrated \HI\ map
  derived from the $30''$ VLA data from \citet{ff02}. Grayscale runs
  from $2\cdot 10^{19}$ (white) to $2\cdot 10^{21}$ cm$^{-2}$
  (black). Contours are as in top-left panel. The VLA beam is
  indicated in the bottom-right corner. {\bf Bottom-left:} Intensity-weighted mean velocity
  field derived from the GBT data. The thick black contour indicates
  $v=130$ km s$^{-1}$. White contours decrease with respect to this
  value in steps of 20 km s$^{-1}$. Black contours increase in steps
  of 20 km s$^{-1}$.  {\bf Bottom-right:} Intensity-weighted mean velocity field derived from
  the $30''$ VLA data. Contours are as in bottom-left
  panel.\label{fig:momnew}}
\end{figure*}

Using the clipped data cube described above we create moment maps
showing the integrated emission (zeroth moment) and the
intensity-weighted average velocity (first moment). These are shown in
the left-hand panels in Fig.~\ref{fig:momnew}. In this Figure, we also
show for comparison (in the right-hand panels) the integrated \HI\ map
and velocity field derived from the $30''$ VLA data set from
\citet{ff02}.  The \HI\ distribution in the GBT data, even when taking
the different beam sizes into account, extends further out than in the
VLA data.  Part of this is due to the size of the primary beam of the
VLA.  Figure~\ref{fig:momslice} shows the 50 percent and 10 percent
sensitivity VLA primary beam sizes overplotted on the VLA integrated
\HI\ map. The NGC 2403 emission as observed by the VLA extends out to
the edge of the 50 percent sensitivity primary beam. Beyond this
radius the primary beam sensitivity drops rapidly, hampering the
detection any emission that may be present there. This emission is
however detected by the GBT observations which do not suffer from
primary beam effects.

The velocity field derived from the GBT data cube looks regular, and
shows signs of a slight warp in the outer parts, as indicated by a
small variation of the position angle of the kinematical major axis.

In addition to the full data cube used to create the moment maps
discussed above, we also created a cube containing only the emission
from the cloud. This was done by applying an intensity cut to remove
the very bright emission from the main disk.  We created an integrated
\HI\ map of the cloud, which is shown in the top panels of
Fig.~\ref{fig:momnew} overlaid on the GBT and VLA maps.  The cloud
clearly overlaps with the main disk as seen at the GBT resolution.
The high contrast between cloud and disk emission, combined with the
relatively large size of the beam, means we can only identify cloud
emission where (in the channel maps) it does not overlap with the main disk (see, e.g.,
Fig.~\ref{fig:chans} at $v=124.8$ km s$^{-1}$ where the cloud emission
merges in projection with the main disk emission).  It is therefore
likely that the cloud extends further toward the SE than
Fig.~\ref{fig:chans} suggests, and that the SE edge is artificial. Due
to the confusion with Galactic emission it is not possible to
establish whether other low column density features are present in the
channels with $v<10$ km s$^{-1}$. If present, they would however not
be related to the cloud because of the different spatial location of
the main disk \HI\ at these velocities.

The total \HI\ mass of the cloud is $6.3 \cdot 10^6$ \msun, or about
0.15 percent of the total \HI\ mass of NGC 2403. The radial distance
of the highest column density part of the cloud to the center of NGC
2403 is $\sim 17'$ (corresponding to $\sim$ 16 kpc or $\sim 2$ times
$R_{25}$). The \HI\ mass is a lower limit because, as noted above, we
can only identify cloud emission when it is not projected against the
main disk.

\subsection{Comparison with deep VLA observations}

We compare the GBT data with the deep, single-pointing VLA \HI\ study of
NGC 2403 by \citet{ff01,ff02} as described in Sect.\ \ref{sec:vladata}.

One of the main results to come out of that study is the detection of
an extended, kinematically anomalous \HI\ component in NGC 2403, which
\citet{ff01,ff02} interpret as a thicker disk which rotates slower
than the thin, main, cold \HI\ disk.  The velocity field of the thick
component shows evidence for inflow motion towards the center. In
addition, \citet{ff01,ff02} report the presence of gas at
``forbidden'' velocities, that is, having velocities that make it
impossible for it to be part of the regular rotation of the disk.

In addition, NGC 2403 was found to contain a number of kinematically
anomalous \HI\ complexes that are not at forbidden velocities, but
nevertheless deviate from regular, circular rotation.
\citet{ff01,ff02} interpret the anomalous gas as a sign of accretion.

One of the most prominent of these anomolous \HI\ complexes is a
filament with an \HI\ mass of $\sim 1 \cdot 10^7$ \msun, located in
the inner disk, with a velocity that differs 60--100 km s$^{-1}$ from
the local rotation, and which is referred to as the ``8-kpc filament''
by \citet{ff02}.

\citet{ff02} fit Gaussians to the peaks of the \HI\ profiles and
subtract these from the data to isolate emission with anomalous
velocities. We also use this method to separate out the emission from
the 8-kpc filament in the VLA data cube. We show an integrated
\HI\ map of this filament superimposed on that of the main disk in the
top panels of Fig.~\ref{fig:momnew} (see also Figs.~7 and 8 in
\citealt{sancisi08}). These figures immediately show that the 8-kpc
filament points towards the cloud seen by the GBT.

\begin{figure*}
\centering{\includegraphics[width=0.6\textwidth]{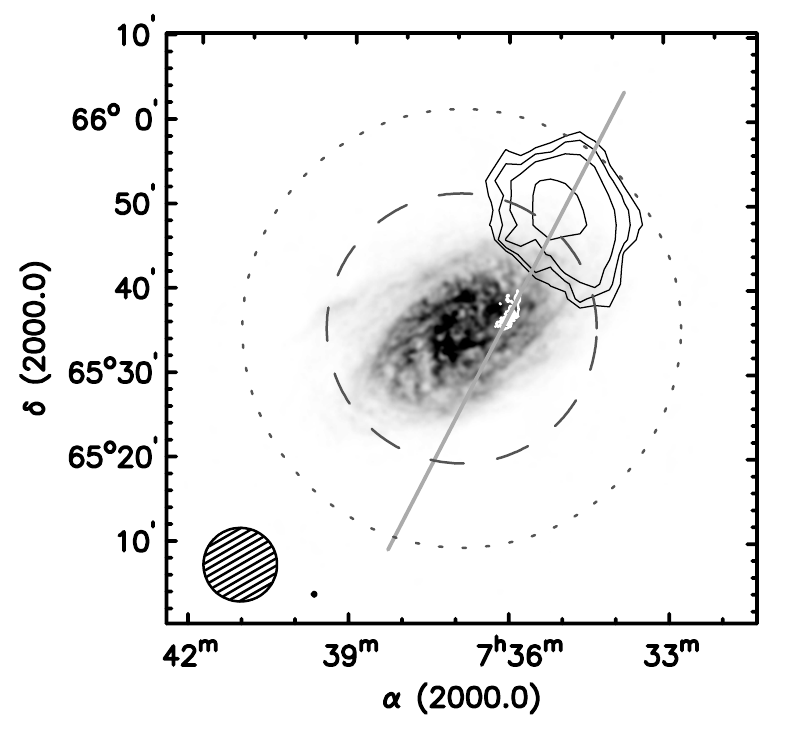}}
\caption{A view of the cloud-filament complex. Grayscales show the
    integrated \HI\ distribution derived from the $30''$ VLA data and
    run from $2\cdot 10^{19}$ (white) to $2\cdot 10^{21}$ cm$^{-2}$
    (black). Thin, black contours show the \HI\ distribution of the
    cloud derived from the GBT data. Contour levels are $(6.25,\,
    12.5,\, 25,\, 62.5) \cdot 10^{17}$ cm$^{-2}$. White contours show
    the 8-kpc filament, derived from the $30''$ VLA data. The contour
    level is $2 \cdot 10^{19}$ cm$^{-2}$. The highest column density
    found in the filament is $6.5 \cdot 10^{19}$ cm$^{-2}$. The GBT
    and $30''$ VLA beams are indicated in the bottom-left corner. The thick gray
    line show the location of the position-velocity slice shown in
    Fig.\ \ref{fig:slices}.  The long-dashed circle indicates the 50
    percent sensitivity level of the VLA primary beam. The
    short-dashed circle shows the 10 percent level.\label{fig:momslice}}
\end{figure*}

\begin{figure*}
\centerline{\includegraphics[width=0.95\textwidth]{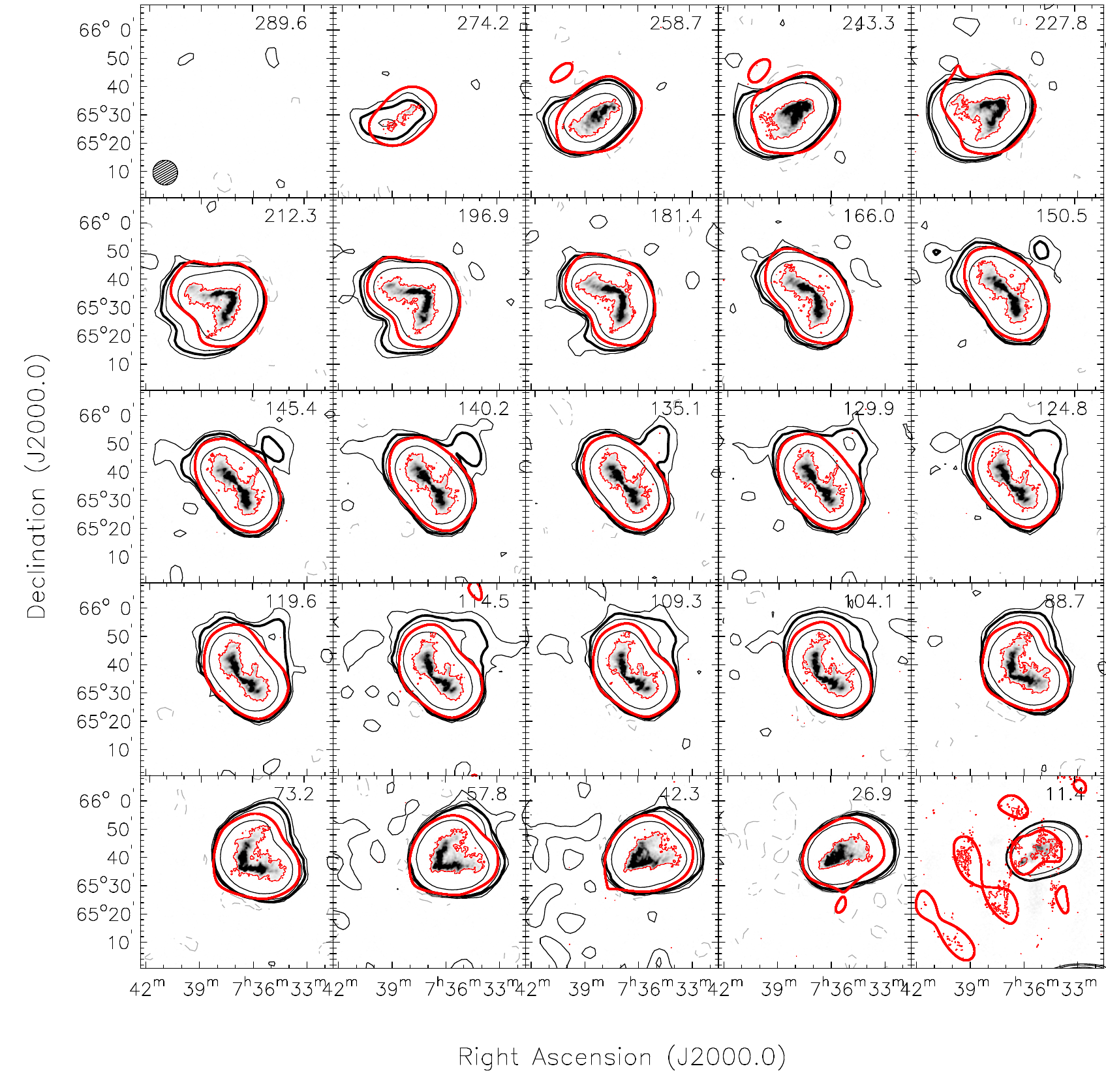}}
\caption{Selected channel maps of the GBT data cube superimposed on
  corresponding channel maps from the deep VLA data.  The GBT data are
  shown in black contours. Black contours levels are $(2.5, 5\ {\rm
    [thick\ contour]}, 10, 50) \cdot 10^{17}$ cm$^{-2}$. Gray, dashed
  contours denote $-2.5 \cdot 10^{17}$ cm$^{-2}$ in the GBT data. The
  gray scale shows the \HI\ emission in the deep $30''$ cube from
  \citet{ff02}. The thin, red contour surronding this emission shows
  the $5 \cdot 10^{18}$ cm$^{-2}$ in these VLA data. The thick, red
  contours show the $5 \cdot 10^{17}$ cm$^{-2}$ level in the VLA data
  when smoothed to the GBT resolution of $8.7'$. This is the same
  column density as shown by the thick black contour. None of the VLA
  contours or grayscales are corrected for primary beam effects. The
  panel in the bottom-right is affected by Galactic emission. The GBT
  beam is shown in the bottom-left of the top-left panel. The two
  small features to the north of the main emission, as indicated by
  the thick red contour at $v=258.7$ and $243.3$ kms s$^{-1}$ are data
  artefacts due to the severe smoothing used.
\label{fig:chansvla}}
\end{figure*}

We make a more detailed comparison in Fig.~\ref{fig:chansvla}, where
we show again the GBT channel maps from Fig.~\ref{fig:chans}, but this
time superimposed on the corresponding deep VLA data channel maps from
\citet{ff02}. Because of the large area shown, the VLA data are not
corrected for primary beam effects.

\begin{figure*}
\centerline{\includegraphics[width=0.9\textwidth]{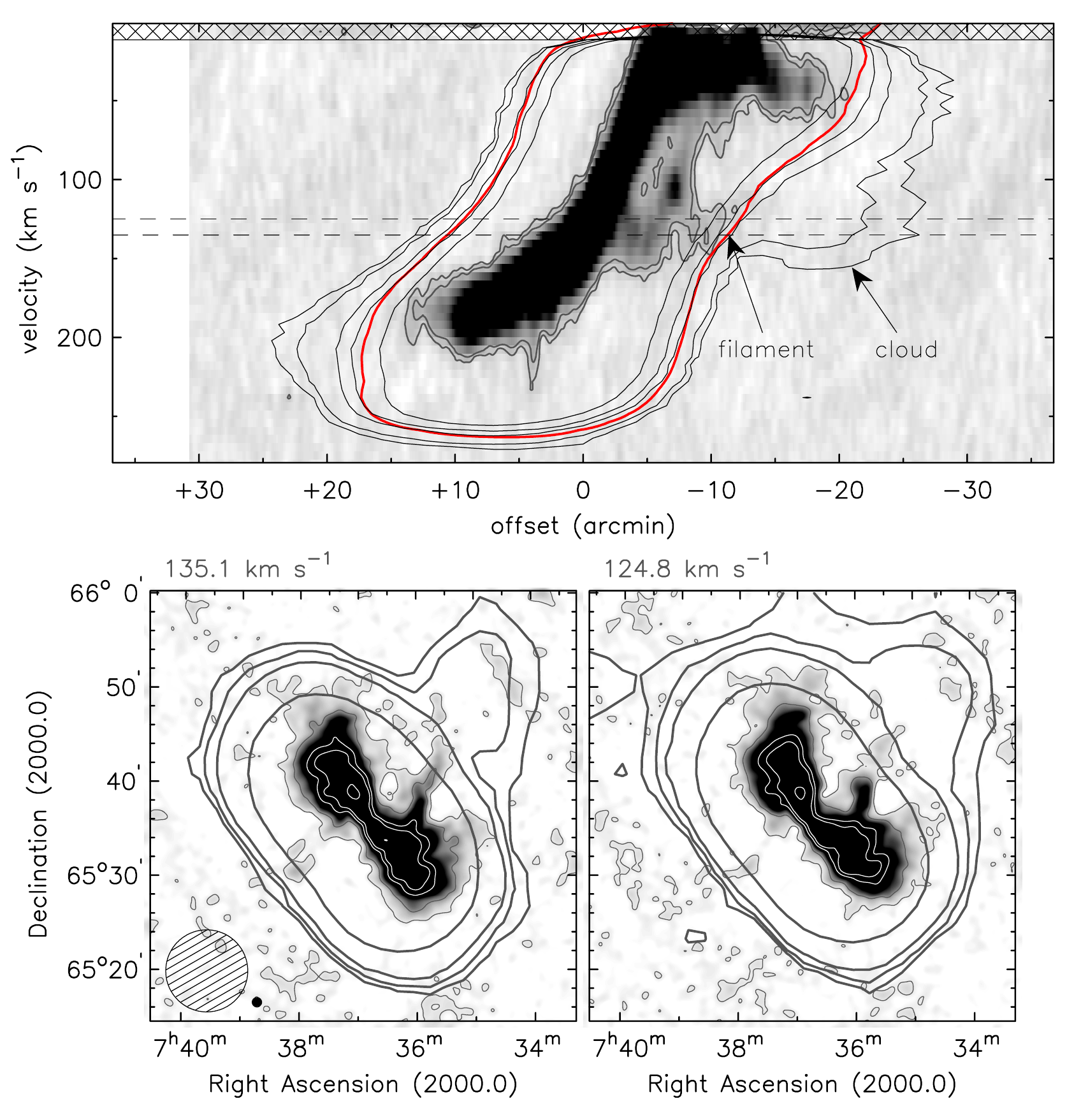}}
\caption{Comparison of GBT and $30''$ VLA data. {\bf Top panel:} $pV$ slice through
  centers of cloud and 8-kpc filament. Position and orientation of the
  slice is indicated in Fig.\ \ref{fig:momslice}. The thickness of the
  slice through the VLA data is $100''$; the thickness of the slice
  through the GBT data is one GBT beam.  The grayscale shows the
  VLA data with $(2.5, 5) \cdot 10^{18}$ cm$^{-2}$ contours
  overplotted in dark-gray.  The black contours show the GBT data at
  $(5, 10, 25, 50, 100)\cdot \sigma$, where $\sigma$ corresponds to
  $3\cdot 10^{17}$ cm$^{-2}$.  The cloud is detected at the
  $>10\sigma$ level. 
  The red contour shows the VLA data smoothed to
  the GBT resolution at the $5 \cdot 10^{17}$ cm$^{-2}$ level. 
  It is
  clear the emission detected by the GBT is more extended. The VLA
  contours and grayscales are not corrected for primary beam
  effects. The hatched area at the top indicates velocities affected
  by Galactic emission. The cloud and 8-kpc filament are indicated by
  arrows. The two horizontal lines indicate the
  velocities of the channel maps shown in the bottom panels. {\bf Bottom
  panels:} Channel maps at $v=135.1$ km s$^{-1}$ (left) and $v=124.8$
  km s$^{-1}$ (right).  In both panels the grayscale shows the VLA
  data at $60''$. Thin dark-gray contours show VLA column density
  levels at $(1, 5) \cdot 10^{18}$ cm$^{-2}$. Thin white contours show
  VLA column density levels at $(25, 50, 250) \cdot 10^{18}$
  cm$^{-2}$. The lowest contour of $1\cdot 10^{18}$ cm$^{-2}$
  corresponds to $\sim 2.5\sigma$. Features in the extreme top-right
  and bottom-left of the channel maps are deconvolution artefacts.
  The thick black contours show GBT column density levels and are as
  in Fig.\ \ref{fig:chansvla}. The GBT beam and the $60''$ VLA beam
  are shown in the bottom-left of the left channel map.
\label{fig:slices}}
\end{figure*}

In the VLA data in Fig.\ \ref{fig:chansvla}, note the ``finger'' of
emission pointing to the NW away from the main emission between 150.5
and 109.3 km s$^{-1}$.  This is the 8-kpc filament.  There is a clear
correlation between the appearance of this ``finger'' in the VLA data
and the cloud emisson in the GBT data, which strongly hints at
  the cloud and filament being associated, and likely forming one single
  complex.  As mentioned earlier, the spatial separation between the
  filament and cloud features is probably artificial, due to our
  inability to detect faint cloud emission projected against the
  bright main disk. We show two channel maps in more detail in the
  bottom panels of Fig.\ \ref{fig:slices}, where we show the GBT
  emission superimposed on the VLA data smoothed to $60''$.  The
  higher sensitivity that we achieve due to the spatial
  smoothing allows us to trace the ``finger'' further out, to the
  point where it touches the cloud emission detected by the GBT.
  Apart from the spatial continuity, there is also a clear velocity
  continuity which we illustrate using the position-velocity ($pV$)
  diagram shown in the top panel of Fig.\ \ref{fig:slices}. The $pV$
  diagram was created by taking a slice through the VLA and GBT cubes,
  going through both the center of the cloud and the filament, as
  indicated by the line in Fig.\ \ref{fig:momslice}. Due to the large
  difference in resolution between the two datasets, we chose
  different thicknesses for the slices. For the GBT data we extracted
  a slice with a thickness of one GBT beam width. To maximize the
  signal in the VLA data, we extracted a slice with a thickness of
  $100''$, corresponding to the average width of the filament.  The
  comparison between the slices in Fig.\ \ref{fig:slices} clearly
  shows that the filament and the cloud occur at the same velocities
  (and \emph{only} those velocities). The clear spatial and velocity
  association strongly suggests that the cloud and the filament form
  one single complex.

The link with the 8-kpc filament definitively rules out the
possibility that the cloud is a Galactic foreground object not related
to NGC 2403.  Such objects are known to exist in this part of the sky
\citep{chyn09}, but they are typically found at velocities of $\sim
-150$ km s$^{-1}$ and are therefore likely to be part of the Milky Way
high-velocity cloud population.

\begin{figure}
\includegraphics[width=0.45\textwidth]{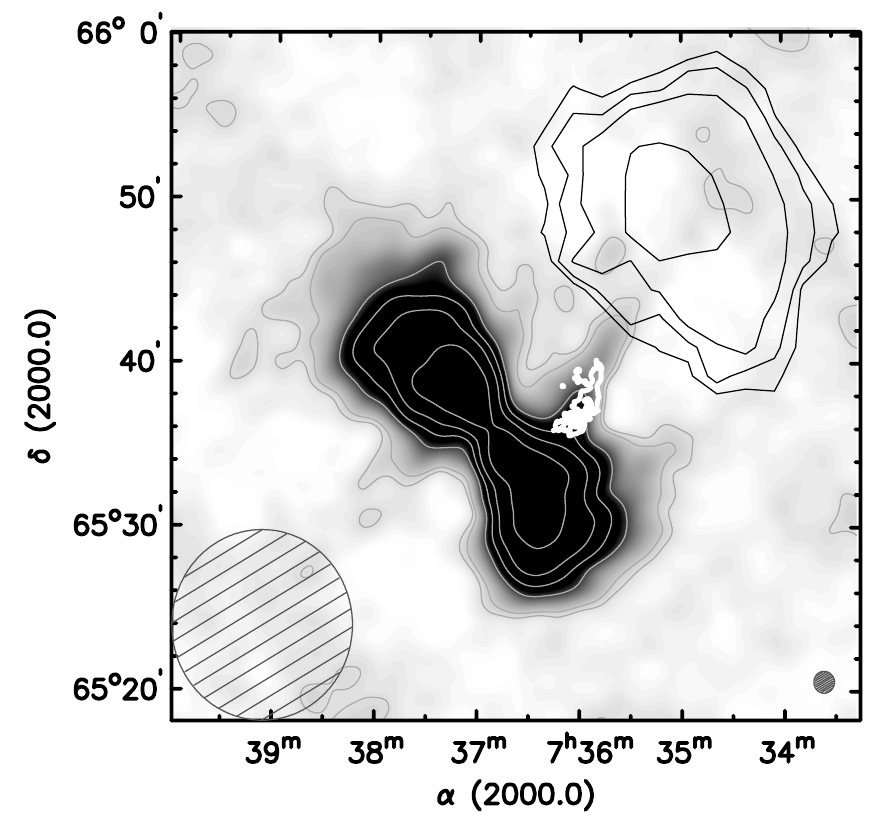}
\caption{Integrated \HI\ map of the emission between 119.6 and 150.4
  km s$^{-1}$ derived from the deep VLA data at $90''$ resolution
  (grayscale). Light-gray contours show $(5,\, 10,\, 20,\, 50,\,
  100,\, 200) \cdot 10^{18}$ cm$^{-2}$ column density
  levels. The noise in the map corresponds to $2 \cdot 10^{18}$ cm$^{-2}$
  and the lowest VLA contour corresponds to $\sim 2.5\sigma$.  Thin
  black contours show the GBT cloud, thick white contours the 8-kpc
  filament at $30''$ resolution. Contours as in
  Fig.\ \ref{fig:momslice}. The $90''$ VLA beam is indicated in the
  bottom-right corner, the GBT beam in the bottom-left corner.
\label{fig:deepmom}}
\end{figure}

We investigated whether the filament could be traced even further out
towards the cloud by convolving the $30''$ VLA data to a beam size of
$90''$.  We created a moment map of all emission between 119.6 and
150.4 km s$^{-1}$ (cf.\ Fig.\ \ref{fig:slices}). No clipping or
masking was applied. The resulting map is shown in
Fig.\ \ref{fig:deepmom}. Compared to the $30''$ data, the emission
extends twice as far and the tip of the filament overlaps with the SE
part of the cloud.  Figure\ \ref{fig:deepmom} shows hints that at the
$\sim 1.5 \sigma$ to $\sim 2 \sigma$ level the filament extends even
further towards the center of the cloud. However, at this level, the
data are also affected by deconvolution artefacts (present as faint
features in the bottom-left and top-right of the Figure) which partly
overlap with the possible faint extension.  In the absence of deeper
high-resolution data it is therefore difficult to say more about this
low-level extension.  Nevertheless, in Fig.\ \ref{fig:deepmom}, the
main filament feature is present at high significance levels and just
overlaps with the (likely artificial) SE of the cloud, further
suggesting that we are dealing with a single cloud/filament complex.

But is the cloud as seen by the GBT indeed a new feature which has not
been seen before, or is it simply the already known 8-kpc filament as
observed with the VLA, but smoothed to a lower resolution?  To answer
this, we use the VLA data cube smoothed to the GBT resolution, as
described in Sect.\ \ref{sec:vladata}.  We compare corresponding
channel maps and a position-velocity slice of this smoothed VLA data
set and the GBT data in Figs.~\ref{fig:chansvla} and \ref{fig:slices},
respectively.  From these Figures it is clear that the cloud emission
was not detected in the single-pointing VLA observations, and that the
GBT observations are showing additional \HI. An explanation for this
was already mentioned: the single-pointing VLA observation has a limited
field of view, and as can be seen in Fig.~\ref{fig:momslice}, the
cloud is located in the region of the primary beam where the VLA
sensitivity has dropped to between 50 and 10 percent of the central
value.

Can we say anything about the column densities of the cloud compared
with that of the filament?  The \HI\ mass of the cloud is $6.3 \cdot
10^6$ \msun. If we take the dimensions of the cloud to be $16' \times
12'$ (i.e., the angular size of the cloud as shown in
Fig.\ \ref{fig:momslice} corrected for the GBT beam size), then the
average \HI\ column density of the cloud is $\sim 6 \cdot 10^{18}$
cm$^{-2}$ assuming a uniform distribution. This would not be detected
by the deep VLA observations.

Due to the GBT beam size we can, however, not distinghuish between
smooth emission which fills the beam, or a more clumpy, higher column
density distribution with a smaller beam filling factor. To take the
other extreme, if we assume that the \HI\ in the cloud has a size and
distribution identical to that of the 8-kpc filament, then the
inferred average column density increases to $\sim 6 \cdot 10^{19}$
cm$^{-2}$. The column densities in the actual observed 8-kpc filament are in
the range $\sim 3\ {\rm to}\ \sim 6 \cdot 10^{19}$ cm$^{-2}$, so
consistent with this value.

It would however still be difficult to detect this more compact
emission in the existing $30''$ VLA data.  Putting the observed 8-kpc
filament at the position of the cloud means applying a primary beam
loss of sensitivity of a factor of $\sim 0.4$. The resulting
\emph{apparent} column density would thus be $\sim 1.5\ {\rm to}\ 2.5
\cdot 10^{19}$ cm$^{-2}$. Assuming that the emission has a velocity
width of $\sim 50$ km s$^{-1}$, implies that in the $30''$ VLA data,
an 8-kpc filament at the position of the cloud would show at the $\sim
1\sigma$ level in each channel (cf.\ the sensitivities listed in
Sec.\ \ref{sec:vladata}).

Figure \ref{fig:deepmom} shows that the $90''$ VLA data has detected more
\HI\ associated with the filament than the $30''$ data.
The ``extra'' emission detected in the former data set (compared to the
$30''$ maps) has a corresponding \HI\ mass between $\sim 3\cdot
10^5\ M_{\odot}$ and $\sim 8\cdot 10^5\ M_{\odot}$, depending on how
exactly the extent of the $30''$ filament is defined. This is only
between 5 and 13 percent of the \HI\ cloud mass detected by the GBT, so it
is clear that even with the extra spatial smoothing applied here, the
VLA does not detect all of the cloud emission.

In an attempt to further constrain the nature of the cloud/filament
complex, we observed the cloud with the WSRT in Aug and Dec 2013 for a
total of 21 hours. The telescope was pointed at the center of the GBT
cloud. The data were reduced using standard methods, and an image cube
was produced using a robustness parameter of 0 and a taper of
$30''$. The resulting beam size was $42'' \times 47''$, with a noise
level of 0.9 mJy beam$^{-1}$. This corresponds to a $5\sigma$, 10 km
s$^{-1}$ column density sensitivity of $2.5 \cdot 10^{19}$ cm$^{-2}$.

The WSRT data confirm all features seen in the deep VLA data, but do
not detect significant emission at the position of the cloud. The
similar limits derived from both the WSRT and the VLA data thus
indicate that the cloud \HI\ must have column densities below $\sim
10^{19}$ cm$^{-2}$ (when observed at $30''$ and integrated over the
velocity width of the cloud) for it not be detected in the deep VLA,
or the more shallow but targeted WSRT observations. The combination of
this column density limit, and the total \HI\ mass of the cloud,
implies that the true spatial distribution of the \HI\ must be fairly
uniform, on scales comparable to the extent of the GBT-detected
emission, as more clumpy \HI\ distributions with higher local column
densities would have been detected.

\section{Interpretation}

\subsection{Comparison with optical data}

We investigate whether any signatures or counterparts of the cloud
are visible at other wavelengths. Inspection of GALEX images
\citep{gilpaz} shows no sign of an overdensity of near-ultraviolet
(NUV) or far-ultraviolet (FUV) sources in the area of the cloud, so
there is no indication of a concentration of recent star formation at
that location.

A study by \citet{barker} looks at the resolved stellar population in
and around NGC 2403, using wide-field observations with Suprime-Cam on
the Subaru telescope.  They used colour-magnitude
diagrams to identify different stellar populations and found that
the stellar distribution of NGC 2403 extends to at least $\sim 40$
kpc, at a surface brightness level of $\mu_V \sim 32$ mag
arcsec$^{-2}$. The field of view of these observations thus
comfortably encompasses the location of the cloud, which is at a
radius of $\sim 17'$ or 16 kpc.

\begin{figure}
\centerline{\includegraphics[width=0.49\textwidth]{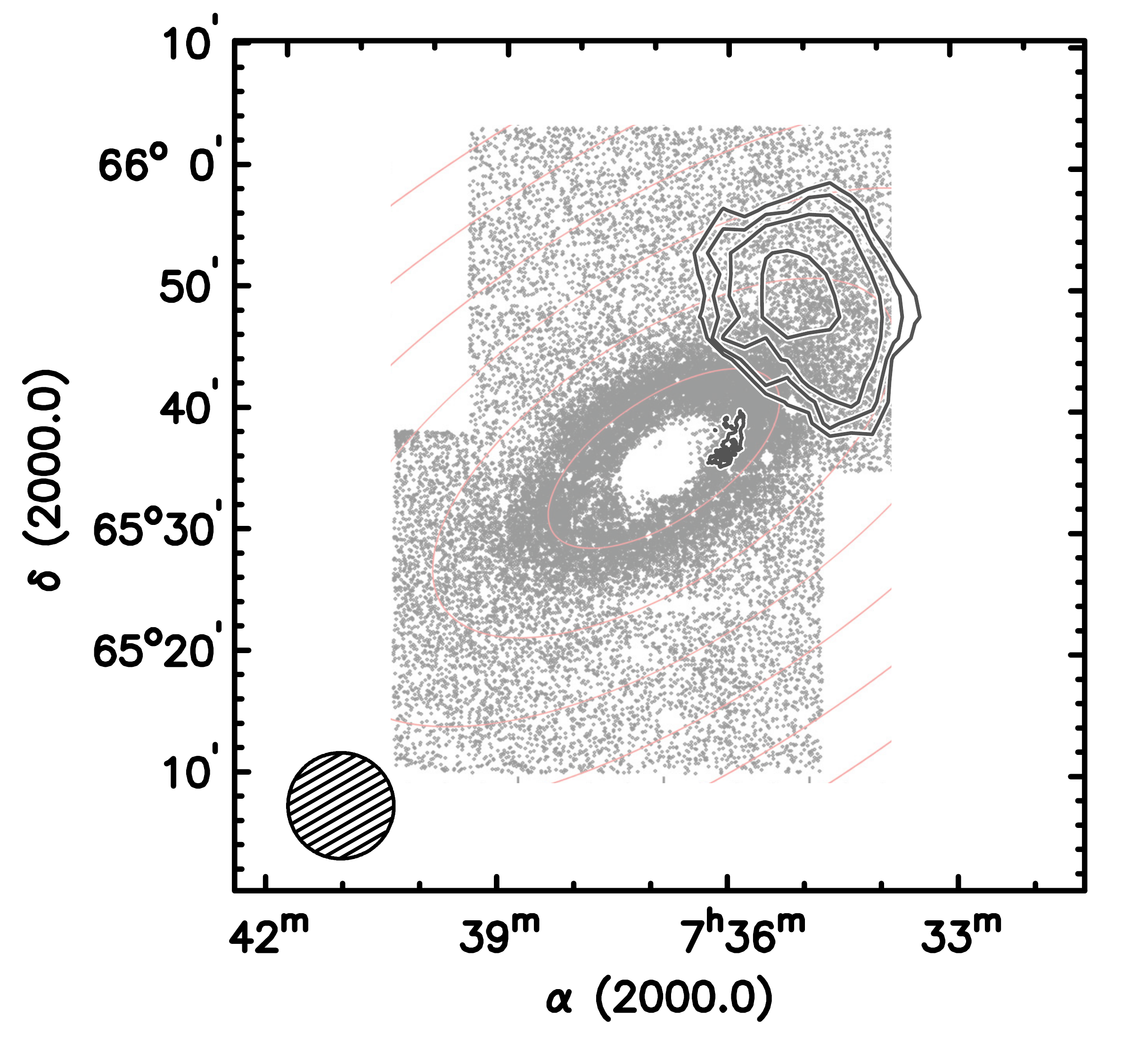}}
\caption{The cloud and filament (contour levels as in
  Fig.~\ref{fig:momnew}) overlaid on the distribution of RGB stars from \citet{barker} 
  (grayscale).  The ellipses are spaced 10 kpc in deprojected
  radius. The absence of stars in the center is due to crowding. The
  GBT beam is shown in the lower left of the panel. The distribution
  of RGB stars is copied from Fig.\ 10 of \citet{barker}.
\label{fig:stars}}
\end{figure}

\citet{barker} present the spatial distribution of the stellar
populations in their Fig.\ 10. Their plot of the distribution of RGB
and AGB stars shows diffuse overdensities of stars to the NW and SE of
the main disk, with the NW overdensity being most prominent.  In
Fig.~\ref{fig:stars} we show the distribution of RGB stars in NGC 2403
from \citet{barker}, and overlay the cloud and the filament. The
center of the cloud is offset $\sim 5'$ to the NE from the approximate
center of the stellar overdensity, but still has a substantial
overlap. 

The optical overdensity shows little internal structure and is likely
part of the faint outer stellar disk, possibly tracing the warp as
seen in \HI\ (cf.\ Fig.\ \ref{fig:momnew}). As noted above, according
to \citet{barker}, stellar emission can be traced out to well beyond
the cloud. The (radially averaged) surface brightness at the radius of
the cloud is $\mu_V \sim 29$ mag arcsec$^{-2}$.

The location of the cloud near the optical overdensity may thus be a
chance superposition and the cloud/filament complex is possibly the
equivalent of a large high velocity cloud (HVC).  The \HI\ mass of the
complex is similar to that of the Complex C HVC in our Galaxy.  This
raises the question of whether the cloud/filament complex could be
caused by galactic fountain processes, i.e., could it be gas blown out
of the main disk due to star formation and forming a HVC? As discussed
above, \cite{ff01,ff02} show the presence of a thick, lagging
\HI\ disk in NGC 2403.  \citet{ff_binney06, ff_binney08} model this
extra-planar gas component using particles that are blown out of the
disk on ballistic orbits. \citet{ff_binney08} conclude that the
general distribution of the gas can indeed be modelled in this way,
but that smaller substructures and large filaments (like the 8-kpc
filament) are not easily produced in the fountain model. With the
addition of the GBT data, the 8-kpc filament is now twice as massive
and extends much further out than originally assumed, making it less
likely to be purely the result of the galactic fountain process.

\subsection{Comparison with NGC 891: a fly-by scenario}

We already noted that another galaxy in which extraplanar gas has been
observed and extensively studied is NGC 891. A paper by
\citet{mapelli}, which focusses on the origin of the lopsided disk in
NGC 891, offers an interesting alternative explanation for the
presence of some of the extra-planar gas observed in that galaxy.

\citet{mapelli} mention that the main extra-planar filament in NGC 891
is pointing at nearby LSB galaxy UGC 1807 (as first pointed out in
\citealt{oosterloo07}; cf.\ their Fig.~8), and using simulations, show
that a fly-by interaction with an intruder dwarf galaxy can create
this filament. Note, incidentally, that these types of fly-bys are
also a common phenomenon in cosmological cold dark matter simulations
\citep{cdmflyby}.

For the specific case of NGC 891, \citet{mapelli} find the best match with an
interaction that happened 300 Myr ago, at a fly-by velocity of 260
km s$^{-1}$. They show that the filament must consist mostly of
gas stripped off the UGC 1807 intruder.

The \citet{mapelli} simulations indicate the presence of gas between
the two galaxies.  Deep GBT \HI\ observations of the area around NGC
891 do, however, not detect any connection between NGC 891 and UGC
1807 (D.J. Pisano, priv.\ com.)  implying that if the fly-by
hypothesis holds, then this gas must be of very low column density
and/or ionized, and only close to the NGC 891 disk dense enough to be
detected.

\begin{figure}
\centerline{\includegraphics[width=0.49\textwidth]{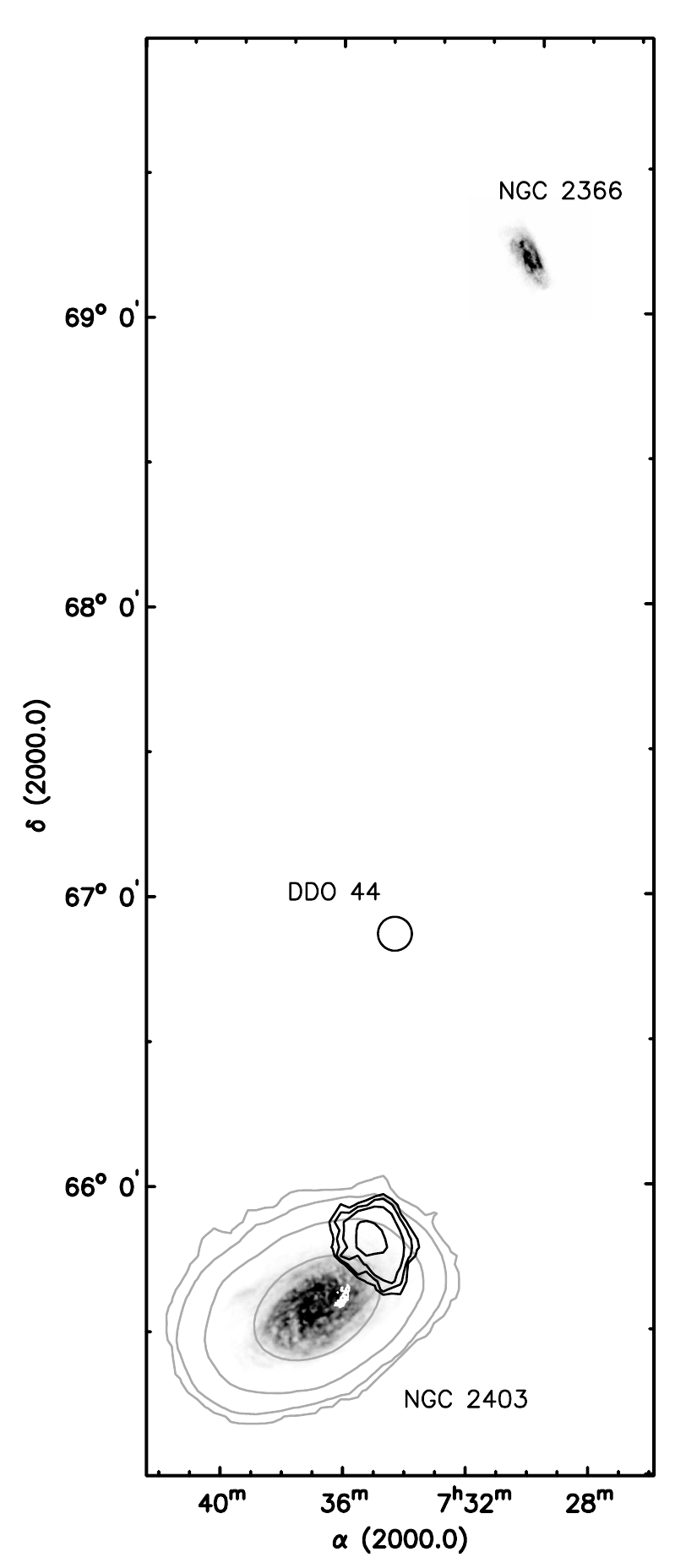}}
\caption{The field around NGC 2403. Contours and grayscales of NGC
  2403 (bottom of the plot) are as in Fig.~\ref{fig:momnew}. The circle
  indicates the position and approximate size of DDO 44.  In the top
  of the panel the \HI\ distribution of NGC 2366 from THINGS \citep{walter08} is
  shown. All distances and sizes are to scale. \label{fig:field}}
\end{figure}

Could the cloud-complex in NGC 2403 be caused by a similar fly-by of a
neigboring galaxy? If we extend the major axis of the VLA 8-kpc
filament\footnote{We do not include the GBT cloud as its morphology is
  too uncertain.} (using a major axis position angle of $-17^{\circ}$;
cf.\ Fig.\ \ref{fig:momslice}) we find that it passes close to the
dwarf spheroidal galaxy DDO 44 (at about 1.5$^\circ$ or $\sim 85$ kpc
from NGC 2403) and also the late-type dwarf galaxy NGC 2366 (at $\sim
3.5^{\circ}$ or $\sim 190$ kpc).  The velocities of the
\HI\ in NGC 2366 completely overlap with those of the \HI\ in NGC
2403.  In NGC 2403, \HI\ is found from $\sim -5$ to $\sim 270$ km
s$^{-1}$. In NGC 2366, this ranges from $\sim 40$ to $\sim 170$ km
s$^{-1}$ \citep{walter08}.  DDO 44 has a single radial velocity
determination that puts it at 213 km s$^{-1}$ \citep{kar11}, which is
inside the range of velocities found in NGC 2403.  See
Fig.~\ref{fig:field} for an overview.

Arbitrarily assuming the same fly-by velocity of 260 km s$^{-1}$ as
given in \citet{mapelli} (and assuming that this velocity is in the
plane of the sky, ignoring any radial velocities), we find the
timescale for an interaction with DDO 44 to be $\sim 300$ Myr. For NGC
2366 we find $\sim 750$ Myr. The DDO 44 timescale is comparable to the
one \citet{mapelli} found for NGC 891. 

Given the orbital time of NGC 2403, which is $\sim 700$ Myr at the
approximate position of the cloud, it is not clear which part of the
stellar disk would have been affected most.  However, the effects of
the interaction on anything other than the local gas distribution must
have been minor.  In a study of the star formation histories at
various locations in the NGC 2403 disk, \citet{williams} find no
evidence for any major temporal variations in the star formation rate.
The locations studied were all on the receding side of the galaxy
(i.e., the side of the galaxy \emph{opposite} to that of the
cloud). Given the regularity and the symmetry of the stellar disk of
NGC 2403, there is, however, no reason to assume that the star
formation history of the approaching side of the galaxy is
dramatically different.  \citet{williams} suggest that as NGC 2403 is
so undisturbed, it is unlikely to have had major interactions in the
past.

Do any of the two potential intruder galaxies show any signs of
interactions or disturbance? NGC 2366 is a gas-rich dwarf-galaxy which
is known to have a disturbed \HI\ velocity field and an outer gas
component that does not follow regular rotation \citep{deblok08,oh08}.
It is undergoing a star-burst and shows evidence for major gas
outflows \citep{eymeren09}. \citet{mcquinn}, from modeling of the
stellar population, find that the star formation burst started and
peaked $\sim 450$ Myr ago.

The second, closer, galaxy DDO 44 is a dwarf-spheroidal galaxy.  In a
search for H$\alpha$ clumps in the M81 group, \citet{kar11} detected
an H{\,\sc ii} region in this dSph galaxy, containing a number of
blue, late-type B-stars. \citet{kar11} note that, despite the absence
of early-type B- or O-stars, the FUV luminosity of $\sim 1 \cdot
10^{36}$ erg s$^{-1}$ of the detected stars is sufficient to ionise
the H{\,\sc ii} region.  These stars are also seen in GALEX data. The
implied star formation rate (SFR) is $\sim 10^{-5}\ M_{\odot}$
yr$^{-1}$.  Despite this small SFR value, it is surprising to find
recent star formation in such a gas-poor system. No \HI\ has so far
been detected in DDO 44, and we also find no evidence in our deep GBT
cube (which also covers the position of DDO 44).  Assuming, like
\citet{kar11}, that the velocity width of DDO 44 is $W\le 30$ km
s$^{-1}$, puts the 5$\sigma$ \HI\ mass upper limit derived from our
GBT data at $M_{\rm HI} \sim 8 \cdot 10^4\ M_{\odot}$.

In their paper, \citet{kar11} speculate that the star formation is
caused by accretion of gas from the intergalactic medium. If the
fly-by scenario is valid, this might therefore be star formation due
to gas dragged out of NGC 2403's immediate environment. The life-time
of a late-type B-star (a few hundred Myr) is consistent with the
timescale for the fly-by derived above.  

We did search the GBT data cube for evidence of an \HI\ connection
between NGC 2403 and DDO 44, but did not find any emission beyond what
was already detected in the filament, though one may expect that gas
to be of such low column density that it is ionised. In fact, we do
not see any evidence throughout the entire GBT cube for emission not
associated with NGC 2403 or our Milky Way at the $3\sigma$ level ($3.9
\cdot 10^{17}$ cm$^{-1}$ per 5.2 km s$^{-1}$ channel). 
  Nevertheless, the simulations presented in \citet{mapelli} suggest
  that in the fly-by scenario, the intruder and the target galaxy
  should be connected by a stream of gas. Given the absence of this
  stream in emission in the GBT data, it will be interesting to see if
  signs of it can be detected in absorption. Alternatively, careful
  stacking of spectra in the supposed area of the supposed stream
  could yield a result, though the unknown velocities of the stream
  (needed to shift the spectra to a common velocity prior to stacking)
  could hamper this excercise.

\subsection{Accretion}

Alternatively, the cloud could be a sign of accretion happening. This
could be accretion (or rather infall) of a dwarf galaxy, or bona-fide
accretion from the IGM or cosmic web.

To start with the first possibility, the cloud and filament complex
could represent the tail left behind by a dwarf galaxy as it plunges
into NGC 2403 and is being stripped. Unfortunately, as we are seeing
the tail projected against the bright stellar disk of NGC 2403, it
will be difficult to directly detect the stellar population belonging
to the dwarf galaxy. A careful study of the resolved stellar
population in the region where the 8-kpc filament connects up with the
main disk is probably the only way to directly detect this stellar
population, if present.

The second possibility is accretion of gas directly from the IGM.
Simulations show that a fraction of gas falling from the cosmic web
into a galaxy halo potential does not get heated to the virial
temperature of the halo \citep{keres}. These filaments of cooler gas
can penetrate from the IGM through the hot gas in the halo of a galaxy
and directly deliver gas to the disk. This process is commonly known
as ``cold accretion''. However, the gas involved still has a
temperature of $\sim 10^5$ K, and is therefore not directly comparable
with the cold gas we are observing in the \HI\ line. The cold
accretion process appears to be most efficient at high redshifts
$(z>2)$. At low redshifts it seems (according to the simulations) to
mainly occur in low-mass galaxies with $M_{\rm halo} \la 5 \cdot
10^{11}\ M_{\odot}$ \citep{joung}, potentially including NGC 2403. As
noted, the ``cold'' accreting gas would have temperatures too high to
be observed in the \HI\ line. However, further simulations
\citep{joung} show that in the innermost regions of the halo $(R \la
100$ kpc) some of the gas can cool further and form \HI\ clouds. For a
Milky-Way-type galaxy, simulations show that at $z=0$ the amount of
cold gas present in the halo is $\sim 10^8 M_{\odot}$
\citep{fernandez}. This is an order of magnitude higher than the
\HI\ mass of the NGC 2403 complex, but the final stages of accretion
are complex and difficult to model, and it is not clear whether this
difference is significant.  See also \citet{nelson} for a comparison
of simulations done with the smoothed particle hydrodynamics and with
the moving mesh techniques. In the latter the amount of cold gas and
corresponding cold accretion rates are much lower.

Qualitatively, the density distribution of the filament can be
consistent with an accretion scenario.  The VLA observations show a
dense filament close to the disk, with the GBT data showing it with a
lower density further out.  However, whereas the current observations
are consistent with the infall of an accreting cloud, they do not
conclusively prove that accretion from the IGM is indeed happening.

\section{Conclusions}

Deep \HI\ observations obtained with the GBT show the presence of a
$6.3 \cdot 10^6\ \msun$ cloud near NGC 2403. Comparison with deep VLA
observations by \citet{ff01,ff02} suggest that this cloud is part of a
larger complex that also includes the $\sim 1 \cdot 10^{7}\ \msun$
``8-kpc'' anomalous-velocity \HI\ filament in the inner disk of NGC
2403.

Deep optical observations by \citet{barker} show the presence of a
stellar overdensity near the position of the cloud. It is not
  clear however whether the overdensity and the cloud are associated.
We suggest that the cloud/filament complex could be a direct
observation of accretion, or, that we could be seeing the
after-effects of a minor interaction (fly-by) with a neigboring
galaxy.

If we are seeing accretion happening, assuming NGC 2403 is not a
unique galaxy, future observations of other galaxies to similar column
densities should reveal more of these events. Should these not turn up
in significant numbers, then this would be a strong indication that
features as now seen in NGC 2403 are interaction events. It would
imply that significant \emph{observable} accretion from the cosmic web
is not happening or at least rare at low redshifts. Future, deeper
observations with instruments such as the SKA, or its precursors, are
required to be able to routinely probe the environment around galaxies
and quantify the importance of accretion from and the relation with
the cosmic web.

\begin{acknowledgements}
We thank the anonymous referee for constructive comments.
We thank the staff at the GBT for their assistance with the observations.
WJGdB was supported by the European Commission (grant FP7-PEOPLE-2012-CIG \#333939).
The work of DJP was partially supported by NSF CAREER grant AST-1149491.  
FF acknowledges financial support from PRIN MIUR 2010-2011, prot.\ 2010LY5N2T.
\end{acknowledgements}

\end{document}